\documentstyle[prl,aps,multicol]{revtex}
\begin{document}
\tightenlines
\input{psfig}
\newcommand{\be}{\begin{equation}}
\newcommand{\ee}{\end{equation}}
\newcommand{\bq}{\begin{eqnarray}}
\newcommand{\eq}{\end{eqnarray}}
\title{Self-trapping mechanisms in the dynamics of three coupled
Bose-Einstein condensates}
\author{Roberto Franzosi$^*$ and Vittorio Penna$^{\dagger}$}
\address{$^*$ Dipartimento di Fisica dell'Universit\`a di Pisa, 
 and INFN, Sezione di Pisa,  Via Buonarroti
 2, I-56127 Pisa, Italy. }
\address{$^{\dagger}$ Dipartimento di Fisica, Politecnico di Torino,
 and INFM, UdR Torino,
C.$^{so}$ Duca degli Abruzzi 24,
I-10129 Torino, Italy.}
\date{\today}
\maketitle
\begin{abstract}
We formulate the dynamics of three coupled
Bose-Einstein condensates
within a semiclassical scenario based on
the standard boson coherent states.
We compare such a picture with that of Ref. \cite{NHMM}
and show how our approach entails a simple formulation
of the dimeric regime therein studied. 
This allows to
recognize the parameters that govern the 
bifurcation mechanism causing self-trapping, and paves the way 
to the construction of analytic solutions.
We present the results of a numerical simulation showing  
how the three-well dynamics has, in general, a cahotic 
behavior.
\end{abstract}
\begin{multicols}{2}
%
%
An increasing interest for the dynamics of coupled bosonic
wells [known in the literature as the {\it dimer}
({\it trimer}) model in case of a pair (triplet) of coupled
wells] has been prompted recently by the
construction of
devices where Bose-Einstein condensates (BEC) interact
through the tunneling effect (see \cite{GEN} and references
therein).
The theoretic work focused on
such models, both in the atomic physics community and in
other areas of theoretical physics, has supplied a large
amount of results disclosing a quite structured
interwell dynamics.

The two-well model (TWM) --used to represent two coupled
BECs in a symmetric double-well potential-- has been
investigated within a picture based on the algebra su(2)
in Refs. \cite{milb}, where, after stemming the model from
the many-body quantum theory of BECs,
the initial state with the atomic population self-trapped
in one well is shown to evolve in delocalized oscillations
involving both the wells.
%
The same model has been studied
previously in Ref. \cite{afko}, both at the quantum level
and from the point of view of the dynamical system theory,
to illustrate the level splitting that characterizes the
dimer spectrum as a manifestation of the orbit bifurcation
in the dimer phase space. 
In Refs. \cite{SFGS} the dynamics of the asymmetric TWM
have been faced within the mean-field formulation
relatively to the $\pi$-phase oscillations as well as the
self-trapping effect. The latter was considered as
well in Ref. \cite{afko} and therein interpreted as a
symmetry breaking phenomenon.
More recently, the TWM (and its S-well
generalization) has been related
\cite{FPZ,FP} to
the Bose-Hubbard model~\cite{com1} and the two-well
ground-states have been interpreted as
insulator/superconducting regimes.
In particular, reformulating the TWM in
an effective single-boson realization --generalizable to 
any S-well system-- has been
shown to favour the use of the system symmetries
as well as the recognition of the inner parameters
controlling the occurrence of doublets in the energy
spectrum.

In this paper we consider some recent results proving
the existence of
configurations with self-trapping within the dynamics of
symmetric trimer (identical interwell couplings). 
These have been obtained in Ref. \cite{NHMM}
by recasting the trimer Hamiltonian within a two-boson
operators picture (introduced in the sequel)
which involves the algebra su(3).
Such a picture
is the extension of the dimer case~\cite{milb} based on
the su(2) (the formal setup for S-well
models involves~\cite{FPZ,FP} the algebra su(S)).
The main contribution of this paper is to apply 
to the trimer an alternative approach that both
reproduces the results of the su(3) picture
and show how the dynamical mechanism
causing self-trapping not only depends on the 
tunneling amplitude but also from the system initial conditions. 
Such an approach relies on
a boson coherent state formulation previously developed
for both boson and spin lattice models~\cite{AP} which
seems to be very simple and effective. The symmetric trimer
is described by
Hamiltonian
$$
H= U \sum^3_{i=1}\, n^2_{i} -v N -
{\frac{1}{2}} \sum_{i<\ell} T_{i \ell} \,
\left (a^{+}_{i} a_{\ell} + a^+_{\ell} a_{i}  \right ) ,
$$
with $T_{i \ell} = T$, that one
can derive from the many-body quantum theory
of BECs through a three-mode expansion of the
condensate field operator \cite{NHMM}. Parameters
$U$, $v$, $T$, account for the interatomic scattering,
the external potential and the tunneling amplitude,
respectively; $n_{i}\doteq  a^{+}_{i} a_{i}$
count the bosons in the $i$th well ($N=\Sigma_i n_i$), 
while the destruction
(creation) operators $a_{i}$ ($a^{+}_{i}$)
obey the canonical commutators
$[a_{i},a^{+}_{\ell}]=\delta_{i\ell}$.
Preceding studies of the trimer dynamics have been
focused on the asymmetric case characterized by tunneling
amplitudes $T_{12} \gg T_{13}, T_{23}$.
Classically ($a_{i} a^{+}_{\ell}
= a^{+}_{\ell} a_{i}$, $a^{+}_{\ell} \equiv
a_{\ell}^{*}$), the asymmetric trimer
has revealed~\cite{hg} the presence of homoclinic chaos,
while, at the quantum level, the survival of breather
configurations~\cite{ff} has been investigated on the
trimer viewed as the smallest possible closed chain.

If one derives the Heisenberg equations related to $H$
for the boson operators $a_{i}$, $ a^{+}_{i}$
and implements the random phase approximation in the
equations for their expectation values
$z_i= \langle  a_{i} \rangle$,
$z_i^* = \langle a^+_{i} \rangle$,
the resulting equations
for the three-well dynamics are $(j=1,2,3)$
\be
i\hbar {\dot z}_j =(2U|z_j|^2 -v+T/2) z_j
- T(z_1+z_2+z_3)/2
\, , 
\label{EM3}
\ee
which entail $\Sigma_i |z_i|^2$ as a conserved quantity
replacing the total boson number $N$ such that
$[N, H]=0$.
%
The Hamiltonian structure of the Heisenberg equations
is inherited by Eqs. (\ref{EM3}) that, in fact, are
also obtained from 
$ {\cal H}( Z, Z^*)$ 
$\equiv \Sigma^3_{j=1}
[ (U |z_j|^2 -v )|z_j|^2 -
T (z^{*}_j z_{j+1} + c.c. )/2
] ,
$
by using the standard Poisson brackets
$\{z^*_k , z_j \}= i \delta_{kj}/\hbar$.

Another significant way to obtain Eqs. (\ref{EM3})
from $H$ relies
on applying the time-dependent variational principle
on a suitable trial state
$|\Psi \rangle = e^{i \theta} |Z \rangle$
with $Z=(z_1, z_2 ,...)$, where $z_r$'s are time-dependent complex
parameters accounting for the system evolution.
Performing the variation of
$\langle \Psi | (i \partial_t -H) |\Psi \rangle=0 $
furnishes a system of hamiltonian equations for
$Z= (z_1, ..., z_r)$ and identifies $\theta$ with the
action of the system. If the trial state is defined
as~\cite{AP}
\be
|\Psi \rangle = e^{i \theta}
|z_1 \rangle \otimes |z_2 \rangle \otimes|z_3  \rangle \, ,
\ee
where $|z_i \rangle$ are the standard bosonic coherent states
that obey the defining equation
$a_{i} |z_i \rangle = z_i |z_i \rangle $,
then Eqs. (\ref{EM3}) are recoverd (up to the shift
$v \to v+U$ ) in which
$z_i \equiv \langle z_j | a_{i} |z_i \rangle $,
$z_i^* \equiv \langle z_j | a^+_{i} |z_i \rangle = z^*_i$,
$|z_i|^2 \equiv \langle z_j | n_{i} |z_i \rangle $,
and $d\theta/dt$ is the Lagrangian associated to $\cal H$.
In addition to describing the system evolution through
$|\Psi \rangle$, this approach also provides a
natural way to find the quantum configuration (in terms of
states) corresponding to the initial conditions of
a given classical motion.

In Ref. \cite{NHMM} the semiclassical treatment of
the trimer dynamics was based on deriving the equations
of motion for the expectation values of the two-boson
operators forming the basis of su(3) instead of
$a_{i}$, $a^{+}_{i}$. Such an algebra
is generated by the creation operators
$
\epsilon_1 = a_1^+ a_2$, 
$\epsilon_2 = a_2^+ a_3$, 
$\epsilon_3 = a_3^+ a_1$,
the destruction operators $\epsilon^+_i = (\epsilon_i)^+$,
$i = 1,2,3$ and the (so-called) Cartan operators 
$h_2 = (D_2-D_3)/\sqrt{3}$, $h_1 = D_1$, where
\be
D_1 = \frac{n_1 - n_2}{2} \, ,\;
D_2 = \frac{n_2 - n_3}{2} \, ,\;
D_3 = \frac{n_3 - n_1}{2} \, .
\label{cart}
\ee
By using {\it imbalance} operators (\ref{cart}),
the su(3) algebraic structure is specified
by the commutators
$[\epsilon_i, \epsilon^+_i]=2D_i$, 
$[\epsilon_i, \epsilon_{\ell}]=
\varepsilon_{i {\ell} k}\, \epsilon^+_k$,
$[D_{\ell}, \epsilon_{\ell} ]= \epsilon_{\ell}$
with $i, k, \ell \in [1,3]$ ($\varepsilon_{i {\ell} k}$ is
the standard antisymmetric symbol),
together with $[\epsilon_i, D_{\ell}]= \epsilon_i /2 $,
and $[\epsilon_i, \epsilon^+_{\ell}] = 0 $,
for $i \ne {\ell}$.
%
Expressing Hamiltonian $H$ through 
$h_1$ and $h_2$ one finds
\be
\!\!\!\!\!
H = 2U( h_1^2 + h_2^2)-f(N) -
\frac{T}{2} (\epsilon_{1}+\epsilon_{2}+\epsilon_{3} + h.c.)\, ,
\label{HJ3}
\ee
with $f(N):= UN^2/3 +vN$, where the operator
$N$ is a group invariant, namely $[N, g]=0$,
$\forall g \in$ su(3). This implies that $[N, H]=0$.
In this framework the Heisenberg equations are
easily carried out. If the random phase approximation
$\langle AB \rangle \equiv \langle A \rangle
\langle B \rangle $ is also implemented
Heisenberg's equations for the su(3) generators
take the form
\be
\cases{
&$
i{\dot {\epsilon_k}} = -(T+4U {\epsilon_k})\, D_{k} +
\, \frac{T}{4}
\varepsilon_{k i \ell} (\epsilon^+_{i} -\epsilon^+_{\ell})\, ,
$\cr
&${\-}$ \cr
&$
i{\dot {h_1}}
= \frac{T}{4}
[(2\epsilon_{1} -\epsilon_{3}-\epsilon_{2})\, -\, c.c. ]\, ,
$\cr
&${\-}$ \cr
&$
i{\dot {h_2}} = \frac{T}{4}
[\sqrt{3} (\epsilon_{3} -\epsilon_{2})\, - \, c.c.] \, . $\cr}
\label{EQH}
\ee
where we have used the displacement operators
$D_j$ for simplifying the formulas.
Notice that, in Eqs. (\ref{EQH})
the approximation
$\langle AB+BA \rangle  \equiv
2 \langle AB \rangle $ has been repeatedly
applied to bilinear terms, and
$\epsilon_{\ell}$, $\epsilon^+_{\ell}$, $h_1$, $h_2$
have been used in place of their expectation values
$\langle \epsilon_{\ell} \rangle$,
$\langle \epsilon^+_{\ell} \rangle$,
$\langle h_1 \rangle$, $\langle h_2 \rangle$.
A possible integrable regime is achieved by setting
$$
h_1 = 0 \; (\Leftrightarrow n_1 \equiv n_2) \, ,\quad
\epsilon_{2} -\epsilon^+_{3}=0 \, ,
\quad \epsilon_{1} -\epsilon^+_{1}=0 \, ,$$
which leads to the reduced system of equations
\be
\cases{
&$ i{\dot {\epsilon_1}} = \frac{T}{2}
(\epsilon^+_{2} -\epsilon_{2} )
$\cr
&${\-}$ \cr
&$ i{\dot {\epsilon_2}} = \frac{T}{2}
(\epsilon_{2} -\epsilon^+_{1} -2D_{2})
-4U \, D_2 \, {\epsilon_2} $\cr
&${\-}$ \cr
&$ i{\dot {h_2}}
= \frac{T}{2}
[\sqrt{3} \epsilon^+_{2}\, - \, c.c.] \; .
$\cr}
\label{RED}
\ee
Their solutions have been calculated implicitly
by geometric arguments and reproduced numerically
for various choice of initial conditions
in Ref. \cite{NHMM}.

In the alternative solution scheme based on Eqs. (\ref{EM3})
the above constraints reduce to impose the condition
$z_1 = z_2$. This selects an integrable sub-dynamics.
In fact, Eqs. (\ref{EM3}) become two,
\be
\cases{
&$ i\hbar {\dot z}_1 =(2U|z_1|^2 -v) z_1
-\frac{T}{2} (z_1 +z_3) $\cr
&${\-}$ \cr
&$ i\hbar {\dot z}_3 =(2U|z_3|^2 -v) z_3 -T z_1\; , $\cr}
\label{EM4}
\ee
where the two costants of motion corresponding to the energy
and the total boson number (we set $n_i \equiv |z_i|^2$)
\be
\!\!
\cases{
&$ \!\!\!\!\! E= U(2 n^2_{1}+ n^2_{3}) -vN -T n_{1} -
T ( z^*_3 z_1 + z^*_1 z_3 ) $\cr
&${\-}$ \cr
&$ \!\!\!\!\! N= 2n_1 \, +\, n_3 $\cr}
\label{COST}
\ee
make Eqs. (\ref{EM4}) integrable.
%
The dynamical behavior is obtained explicitly
via a standard quadrature procedure
(see Refs. \cite{SFGS,MP}) which
furnishes the phase-independent equation
for $D_3$
\be
{\dot D}_3^2 = \frac{9}{16} (4T^2 n_1 n_3 - R^2) 
\label{ESP}
\ee
by substituting
$
R :=[E+vN +T n_{1}-U(2 n^2_{1}+ n^2_{3})]$ $=-T(z_3^* z_1 + C.c.)
$
inside the (squared) equation
${\dot D}_3^2 = -9T^2[z_3^* z_1 - c.c.]^2/16$ for $D_3$.
Introducing the further constant of motion $N$ to obtain
${\dot D}_3^2$ written in terms of the unique variable
$D_3$ requires that $n_1$ and $n_2$ are expressed as
$n_1 = (N- 2D_3)/3$ and $n_3 = (N+ 4D_3)/3$. These,
in turn, substituted in Eq. (\ref{ESP}) give the equation
\be
{\dot D}_3^2 =
\frac{T^2}{4} (N- 2D_3)(N+4 D_3)\, -\frac{9}{16}\, R^2(D_3)
\, ,
\label{EQU}
\ee
for the imbalance variable $D_3= (n_3- n_1)/2$, in which
$$
\!\!\!\!\!\!\!\!\!\!\!\!\!\!\!\!
R(D_3) \equiv E+ vN + \frac{T}{3} (N- 2D_3)
-\frac{U}{3}(N^2+8D_3^2 )=\,
$$
$$
\quad \quad   = \frac{2}{3} \left \{ (A- D_2)[T + 4U(A+D_2)]
-TN\, K (P)
\right \} \,
$$
with $A:= D_3(0)$,
$K (P)
:=\frac{1}{2} [(a+2)^2-9a^2]^{\frac{1}{2}} {\rm cos}\Delta$,
$P:=(a, \Delta)$,
$a= 2A/N$, and $\Delta := \theta_3(0) - \theta_1(0)$.
The second version of $R(D_3)$ is obtained by writing
$E$ in terms of the initial conditions
$D_3(0)$, $\theta_k(0)$. Phases $\theta_j$ are defined by
$
z_k= 
\sqrt{n_k} e^{i\theta_k}
$.
Eq.~(\ref{EQU}) can be cast in the dimensionless
form $(dx/ds)^2 = -2V_{\tau}(x; P)$ with $s := NUt$ and
$$
V_{\tau}(x; P) :=
\frac{1}{2}    
\bigl [ (a- x)(a+ x+ \tau/2)-\tau K \bigr ]^2 -
\frac{\tau^2}{2} (1-x)(1+2x)
$$
where $x := 2 D_3/N$ ($x \in [-1, 1]$),
$\tau := T/NU$.
In view of the fact
that both the squared term in $V_{\tau}$ (namely  $R^2$)
and $(d x/ds)^2$ are nonnegative,
the further condition $(1-x)(1+2x)\ge 0$ must be
accounted for which implies the restriction of the
$x$ range to $-1/2 \le x \le 1$.

The reduction of Eq.~(\ref{EM4}) to Eq.~(\ref{EQU}) allows
one to construct explicit solutions in terms of elliptic
functions by recasting the quartic term via standard transformation
methods \cite{DAV}. This will be enacted elsewhere.
Operationally, our goal --the description of biforcation mechanism
inherent in Eq.~(\ref{EQU})--
can be achieved as well through the equivalent potential problem 
${\cal E}=$ $ \frac{1}{2} (dx/ ds)^2 + V_{\tau}(x; P)$
at ${\cal E}=0$, where parameters $N$, $K$
and $x(0)$ in $V_{\tau}$ are fixed by setting the initial conditions.

With negative $\tau$ and a suitable choice of the other
parameters, $V_{\tau}$ can exhibit an asymmetric double-well.
In general, three solutions are obtained by
annihilating
$$
\frac{dV_{\tau}}{dx}= 2x^3+ \frac{3\tau}{2} x^2+
\frac{1}{4} \Bigl [9\tau^2 -8 \beta_{\tau} (P) \Bigr ]x
-\frac{\tau }{2} \Bigl [\tau +\beta_{\tau} (P) \Bigr ],
$$
%
%
where $\beta (\tau , P) := (a + \tau/4)^2 -\tau (K+\tau/16)$,
that correspond to a maximum of $V_{\tau} (x; P)$ with
two side minima.

In particular, setting $a =1$ reproduces the
conditions under which dynamics was studied in
Ref. \cite{NHMM}
(depleted twin wells, that is $n_3 (0) \equiv 1$), and
leads to the potential
$$
V_{\tau} (x) =
\frac{1}{2} (1- x)^2 \left (x+ \frac{\tau}{2}+1 \right)^2 -
\frac{\tau^2}{2} (1-x)(1+2x)
$$ 
whose maximum is such that $V_{\tau} (x_m) = 0$
with $x_m = 0$ when $\tau = -2/3$. For $\tau > -2/3$
one has $V_{\tau} (x_m) > 0$. The important feature
thus emerging (see Fig.~1) is that, whenever the
potential maximum is nonnegative, $V_{\tau}(x)$
generates two noncommunicating basins with  $V_{\tau}(x) 
\le 0$ (separated by a forbidden interval 
where $V_{\tau} >0$) entailing two independent oscillatory
motions. In each basins the motion has a periodic
character.
This represents the bifurcation effect reminescent
of the behavior manifested by two-well dynamics
\cite{afko,FP}.

What we emphasize here, based on our $z_j$ description,
is that the onset of separated motions can be caused
by varying the other parameters of the problem.
In particular, a high sensitivity is manifested
relatively to the initial phases
incorporated in $\Delta$. Suitable changes of the latter
are capable of switching on the bifurcation mechanism
even for $a \ne 1$.
Such a situation is represented in Fig. 2 for $a= 0.99$
(twin wells almost empty) and $\tau = -2/3$, where various
potential wells are generated by varying ${\rm cos} \Delta$
in $[-1,1]$. For sufficiently low values of
${\rm cos} \Delta$ the presence of the maximum is ensured.
The `opposite' case $a= -0.49$ and $\tau = -1/3$
(corresponding to twin wells almost
half-filled and $n_3(0) \simeq 0$) of Fig.~3
confirms the presence of isolated basins as well as 
the case with a more negative coupling $\tau = -0.7 < -2/3$
and $a= 0.99$.

Decreasing sufficiently the value of $\tau$ 
(Fig. 4 illustrates the case
$\tau = -0.8$ with $a= 0.99$) by keeping
the same range for ${\rm cos} \Delta$ entails
situations where wells never exhibit a local maximum.
This can be
proved analitically in the special case $\tau = -1$
in which the potential becomes
$$
V_{\tau}(x; P) \equiv
\frac{1}{2}    
\Bigl [ (a-1/4 )^2 + K -X^2 \Bigr ]^2 - \frac{9}{16}+ X^2
$$
with $X= x-1/4$, and the stationary points can be calculated
explicitly. One finds a maximum at $x_m = 1/4$ with
$V_{\tau} (x_m) < 0$ so that no bifurcation effect occurs.
The side minima are placed at
$x_{r, \ell} = 1/4 \pm [K-1+ (a-1/4)^2]^{1/2}$. These are
real provided
$K-1+ (a-1/4)^2 \ge 0$ namely if
$$
{\rm cos} \Delta \ge
[1- (a-1/4)^2]/ [(1-a)(1+2a)]^{\frac{1}{2}}\, .
$$

For a generic $\tau$,
the maximum depends on $a$ and
$\Delta$ in a complicated way which makes difficult
the analytic calculation of $V_{\tau} (x_m)$ and of its sign.
Nevertheless, some necessary conditions ensuring its existence can
be obtained explicitly. As suggested by Figs.~1-3,
increasing $\Delta$ with both $\tau$ and $a$ constant
implies that the maximum
at $x = x_m$ and the left minimum at $x = x_{\ell}$ 
reach the (flex) point $x= c$ for critical value $\Delta\equiv\Delta^*$. 
Since the interval
$[ x_{\ell}, x_m]$ where $dV_{\tau}/dx > 0$ vanishes
for $x_{\ell}, x_m \to c$ then
\be
\lim_{\Delta\to\Delta^*}\bigl (dV_{\tau}/dx \bigr )_{x_l,x_m}
=0=  \bigl ( d^2V_{\tau}/dx^2 \bigr )_c \, .
\label{crit}
\ee
The derivation of the roots of $d^2 V_{\tau}/dx^2 =0$
at $x = c$
$$
x_{\pm} =
\frac{\tau}{4} \bigl \{ -1\pm [
8(2 a^2+ a\tau -2K\tau )/(3 \tau^2)- 5 ]^{1/2}
\bigr  \} \, ,
$$
from
$\,
d^2V_{\tau}/{dx^2} =\, 6 x^2+ 3 \tau x
-2 \beta_{\tau}(P)+9\tau^2/4
$ 
allows one to exploit the fact that
the lowest one, $x_-$, is a maximum of $dV_{\tau}/dx$
corresponding to the $V_{\tau}$ flex point at $x= c$.
When
\be
\bigl (dV_{\tau}/dx \bigr )_c \, \equiv
\frac{8\tau}{3^{3/2} |\tau|}\,
[\beta_{\tau} (P) ]^{\frac{3}{2}}
- \tau (\tau +1) \ge 0
\label{cond}
\ee
becomes negative 
the maximum disappears (see, e. g., Figs. 1-3).
The bifurcation condition $V_{\tau}(x) >0$ must be
searched within the parameter space domain 
where $a$, $\Delta$, $\tau$ satisfy formula (\ref{cond}).

The analysis just developed shows that changing
$\Delta$ can modify deeply the system dynamics
and that, in general, the onset of the bifurcation effect is
governed by the complex interplay of all parameters
$a$, $\Delta$, $\tau$.
A complete study of the dynamics requires that one considers 
any possible initial
condition for the dynamics and thus, e. g., the situations
in which $n_1(0) \ne n_2 (0)$, excluded
in the present paper. In this case the nonintegrable
character of the system crops up in a dramatic way as
illustrated in Fig.~5. The systematic analysis of 
fixed points for the symmetric three-well dynamics
and thus the emergence of chaos close to the
hyperbolic points is in progress at this moment. It
will be discussed in a separate paper.



\end{multicols}
%

\newpage

\begin{figure}[htbp]
\label{fig1}
\caption{By varying $\tau$ in $[-0.75, -0.63]$ with $a=1$, $V_\tau(x,P)$ 
generates a second (small) basin on the left (dashed potential 
corresponds to $\tau\simeq-0.66$).}
\end{figure}
\begin{figure}[htbp]
\label{fig2}
\caption{By varying $\Delta$ in $[0,\pi]$ with $a=0.99$, $V_\tau(x,P)$ 
generates a second (small) basin on the left (dashed potential 
corresponds to $\Delta\simeq 1.40$).}
\end{figure}
\begin{figure}[htbp]
\label{fig3}
\caption{Representation~of~bifurcation~mechanism~by~varying~$\Delta \in 
[0,\pi]$ in $V_\tau(x,P)$~with~$a=-0.49$,~$\tau=-1/3$. $~~~~~~$ $~~~~~~~
$$~~~~~~~~$ $
~~~~~~~~~~~~$ $~~~~~~~~~~~~~~~$}
\end{figure}
\begin{figure}[htbp]
\label{fig4}
\caption{$V_\tau(x,P)$~with~$a=0.99$~and~$\tau=0.8$:
~a~sufficiently~negative~$\tau$ involves a single basin
for~any~$\Delta \in [0,\pi]$.$~~~~~~~~~~~~~~~~
~~~~~~$}
\end{figure}
\begin{figure}[htbp]
\label{fig5}
\caption{Poincar\'e section of the $\xi_1-\phi_1$ plane, 
where $\xi_1:= 1-2n_1/N$ and $\phi_1=\theta_2-\theta_1$, obtained
by setting  $n_3\simeq 6.85$; this is derived by numerical integration 
of Eqs. (1) with energy $E\simeq 92.33$, $N=10$, $T=U=1$.}
\end{figure}


\begin{thebibliography}{000}
\bibitem{com1}
The Bose-Hubbard model for a bosonic fluid
on an S-site lattice~\cite{AP} is interpreted as 
S coupled bosonic wells (representing a chain of interacting
condensate in the mean field approximation) by means of the
time-dependent variational principle and the coherent state
method.
\end{thebibliography}
\end{document}